\documentclass[journal]{IEEEtran}
\usepackage{cite}

\ifCLASSINFOpdf
   \usepackage[pdftex]{graphicx}
\else
\fi
\usepackage{amsmath}
\usepackage{bm}
\usepackage{algorithm}
\usepackage{makecell,multirow,diagbox}

\usepackage{algorithmic}

\usepackage{array}
\begin{document}
%
\title{Weighted Encoding Based Image Interpolation With Nonlocal Linear Regression Model}
%
%
%

\author{Junchao~Zhang
\thanks{Junchao Zhang is with School of Aeronautics and Astronautics, Central South University, Changsha 410083, China. (e-mail: junchaozhang@csu.edu.cn)}}

%
%

\markboth{}%
{Zhang: Weighted Encoding Based Image Interpolation With Nonlocal Linear Regression Model}
%



\maketitle

\begin{abstract}
Image interpolation is a special case of image super-resolution, where the low-resolution image is directly down-sampled from its high-resolution counterpart without blurring and noise. Therefore, assumptions adopted in super-resolution models are not valid for image interpolation. To address this problem, we propose a novel image interpolation model based on sparse representation. Two widely used priors including sparsity and nonlocal self-similarity are used as the regularization terms to enhance the stability of interpolation model. Meanwhile, we incorporate the nonlocal linear regression into this model since nonlocal similar patches could provide a better approximation to a given patch.  Moreover, we propose a new approach to learn adaptive sub-dictionary online instead of clustering. For each patch, similar patches are grouped to learn adaptive sub-dictionary, generating a more sparse and accurate  representation. Finally, the weighted encoding is introduced to suppress tailing of fitting residuals in data fidelity. Abundant experimental results demonstrate that our proposed method outperforms several state-of-the-art methods in terms of quantitative measures and visual quality.
\end{abstract}


%
\IEEEpeerreviewmaketitle

\section{Introduction}
%
%
%
%
\IEEEPARstart{I}{mage} interpolation is a fundamental problem in image processing, aiming to reconstruct a high resolution image from its down-sampled observation. It has wide applications in the fields of digital photography, satellite remote sensing, medical imaging and polarization imaging.

The simplest approach to reconstruct high resolution images is based on liner interpolation including Bilinear, Bicubic and Cubic-spline \cite{1,2,3}. However, these methods estimate each missing pixel from its local neighborhood using weighted average and tend to generate artifacts in high-resolution images. To better preserve image details, more complex algorithms based on natural images priors are proposed \cite{4,5,6,7,8,9,10,11,12}. These methods generally produce better interpolation results than liner interpolation methods. NEDI \cite{4} is the representative edge-guided interpolation method. The local covariance coefficients from a low-resolution image are estimated firstly. Then, these coefficients are used to adapt the interpolation at high-resolution images based on the geometric duality. This method is based on the assumption of local stationarity of the covariance. However, this assumption is not completely valid and this method tends to generate artifacts in high-resolution images. The total variation model \cite{9,10,11,12} is based on another prior, i.e. natural images have small first derivatives.

Sparse representation has been successfully applied in the fields of image processing and computer vision \cite{yang2009multifocus,13,14,15,16,17,18,19,20,21,22,liu2014target,fang2015face,fang2017hyperspectral} and it shows promising results. For image interpolation, coupled dictionaries are jointly learned from the low- and high-resolution image patches in SCSR \cite{16}. In this interpolation model, the low-resolution and high-resolution image patch pair share the same sparse representation with respect to their own dictionaries. Therefore, the sparse representation of a low-resolution image patch can be used to generate the corresponding high-resolution image patch. NARM \cite{18} is another image interpolation method based on sparse representation. Nonlocal autoregressive model is embedded in NARM and nonlocal self-similarity is used as regularization term. Besides, image patches are clustered and each class is encoded by an adaptive compact dictionary. Thus,  good interpolated results rely on accurate clustering. However, to cluster image patches accurately is difficulty since the class number and sample number of each class are difficulty to set. Even though experimental results in \cite{18} showed that NARM outperforms several previous methods including SCSR, this method produces spackle noise which is generated by inappropriate clustering.   A new image interpolation scheme is proposed in \cite{22}. Non-local self-similarity assumption is adopted and over-complete dictionary is learned in this sparse model. In recent years, deep convolutional neural networks \cite{23,24,25,26} provides a new strategy for image interpolation.

In this paper, we propose a new image interpolation approach based on sparse representation. Since a given pixel can be well approximated by nonlocal neighbors, we propose the nonlocal linear regression model and incorporate it into the interpolation model. Meanwhile, image sparsity prior and non-local self-similarity prior are adopted to enhance the stability of interpolation model. Current sub-dictionary learned online by clustering is difficulty since the class number and sample number of each class are difficulty to set. Inadequate clustering will reduce the accuracy of sparse representation . To address this problem, we propose a new approach to learn adaptive sub-dictionary online instead of clustering. For each patch, its non-local similar patches are grouped to train the adaptive compact dictionary, generating a more sparse and accurate representation. Moreover, since the distribution of fitting residuals is irregular than Gaussian and it has heavy tails, we introduce weighted encoding into data fidelity to suppress tailing. Abundant benchmark images are used to evaluate the interpolation performance. Experimental results demonstrate that our proposed method outperforms several state-of-the-art methods in terms of quantitative measures and visual quality.

Our model is based on sparse representation, which is similar to NARM \cite{18}, but our work is unlike NARM. Our contributions can be summarized as: (1) We propose a new approach to learn adaptive sub-dictionary. For each patch, its non-local similar patches are grouped to train the adaptive compact dictionary, generating a more sparse and accurate representation. (2) We incorporate  nonlocal linear regression model into interpolation model. The nonlocal linear regression model is differnet with autoregressive model introduced in \cite{18}. Only weighted average is used in the autoregressive model, but our model uses weights and bias, generating a closer approximation to a given pixel. (3) We introduce weighted encoding into data fidelity to suppress tailing of fitting residuals, which ensures that the distribution of fitting residuals is more like Gaussian and the ${{\ell }_{2}}$ norm can be still used in the data fidelity term.

The rest of the paper is organized as : the proposed interpolation model is described in Section II detailedly.  Section III presents experimental results and discussions. Section IV concludes the paper.
\section{Weighted encoding with nonlocal linear regression}
Following the notations in \cite{13}: for an image $\mathbf{x}$,  ${{\mathbf{x}}_{i}}={{\mathbf{R}}_{i}}\mathbf{x}$ denotes the ${{i}^{th}}$ patch vector and ${{\mathbf{R}}_{i}}$ denotes an extracting matrix. Given a dictionary $\mathbf{\Phi }$, the sparse representation of ${{\mathbf{x}}_{i}}$ over dictionary $\mathbf{\Phi }$ is: ${{\mathbf{x}}_{i}}=\mathbf{\Phi }{{\bm{\alpha }}_{i}}$, where ${{\bm{\alpha }}_{i}}$ is the sparse coding coefficient with a few non-zero entries.  ${{\left\| \cdot  \right\|}_{0}}$ denotes the pseudo-norm that counts the number of non-zero entries in a vector.

\subsection{Image interpolation with nonlocal linear regression}

For image interpolation, it is assumed that a low-resolution image ${\bf{y}} \in {R^M}$ is directly down-sampled from its high-resolution image ${\bf{x}} \in {R^N}$, as formulated by Eq. (1):
\begin{equation}
\mathbf{y}=\mathbf{Dx}.
\end{equation}
Where ${\bf{D}} \in {R}^{M \times N}$ is the down-sampled matrix and $N={{l}^{2}}\cdot M$ with the sampling factor $l$ along the horizontal and vertical dimensions. According to sparse representation theory, the image interpolation problem can be transformed to minimize the following model:
\begin{equation}
\underset{\bm{\alpha }}{\mathop{\min }}\,\left\{ \left\| \mathbf{y}-\mathbf{D\Phi}\bm{ \alpha } \right\|_{2}^{2}+\lambda R(\bm{\alpha }) \right\}\begin{matrix}
   {} & s.t. & \mathbf{x}=\bm{\Phi \alpha }  \\
\end{matrix}.
\end{equation}
Where $R(\cdot )$ is the regularization term and $\lambda$ is the regularization parameter. To improve the above model, we incorporate nonlocal linear regression into it. For nature images, the local linear regression model \cite{6,7} is used according to image local redundancy. However, the image local redundancy is inadequate to reconstruct image structures in high precision. Fortunately, nature images often have a rich amount of nonlocal similar patterns. These nonlocal similar patterns may be spatially either close to or far from each other. Therefore, we can establish nonlocal linear regression model using image nonlocal redundancy. 

For each patch ${{\mathbf{x}}_{i}}$ of size $p\times p$, we can get its similar patches in a large enough local window of size $L\times L$. A patch $\mathbf{x}_{i}^{k}$ is selected as a similar patch to ${{\mathbf{x}}_{i}}$ if the Euclidean distance between them is not greater than the preset threshold. In fact, we can get the first $m$ most similar patches, denoted by $\mathbf{X}\text{= }\!\![\!\!\text{ }\mathbf{x}_{i}^{1}\text{,}\mathbf{x}_{i}^{2},\ldots ,\mathbf{x}_{i}^{m}]$. We define the nonlocal linear regression model between ${{\mathbf{x}}_{i}}$ and $\mathbf{X}$ as:
\begin{equation}
{{\mathbf{x}}_{i}}=\left\langle {{\mathbf{X}}^{T}},{{\mathbf{a}}_{i}} \right\rangle +{{b}_{i}}.
\end{equation}
Where ${{\mathbf{a}}_{i}}$ and ${{b}_{i}}$ are the weight vector and bias respectively, and $<\cdot> $ is the inner product. We denote $\mathbf{X}\text{= }\!\![\!\!\text{ }\mathbf{x}_{i}^{1}\text{,}\mathbf{x}_{i}^{2},\ldots ,\mathbf{x}_{i}^{m},\mathbf{1}]$ and $\bm{\omega }_{i}^{T}=[\mathbf{a}_{i}^{T},{{b}_{i}}]$, the Eq. (3) can be rewritten as:
\begin{equation}
{{\mathbf{x}}_{i}}=\mathbf{X}{{\bm{\omega }}_{i}}.
\end{equation}
Because the weight vector ${{\bm{\omega }}_{i}}$ is used to estimate the center pixel ${{x}_{i}}$ of patch ${{\mathbf{x}}_{i}}$, the weight for each dimension should be different. The closer to the center pixel, the greater the weight. Here we give a simple kernel function to determine the weight as:
\begin{equation}
\kappa ({{x}_{i}},{{x}_{j}})=exp\left( -\frac{\left\| d({{x}_{i}},{{x}_{j}}) \right\|_{2}^{2}}{{{\sigma }^{2}}} \right).
\end{equation}
Where ${{x}_{j}}$ represents neighbor pixels in a window of $p\times p$ and $d({{x}_{i}},{{x}_{j}})$ represents the distance between ${{x}_{i}}$ and ${{x}_{j}}$. Thus, the weight vector ${{\bm{\omega }}_{i}}$ can be determined by solving the following regularized minimization problem:
\begin{equation}
{{\widehat{\bm{\omega }}}_{i}}=\arg \underset{{{\bm{\omega }}_{i}}}{\mathop{\min }}\,\left\{ \left\| {{\bm{\kappa }}^{1/2}}\left( {{\mathbf{x}}_{i}}-\mathbf{X}{{\bm{\omega }}_{i}} \right) \right\|_{2}^{2}+\chi \left\| {{\bm{\omega }}_{i}} \right\|_{2}^{2} \right\}.
\end{equation}
where $\bm{\kappa }=diag(\kappa ({{x}_{i}},{{x}_{j}}))$, $\chi $ is the regularization parameter and used to avoid overfitting. The solution of Eq. (6) can be obtained by setting the derivative to zero.
\begin{equation}
{{\widehat{\bm{\omega }}}_{i}}={{\left( {{\mathbf{X}}^{T}}\bm{\kappa X}+\chi \mathbf{I} \right)}^{-1}}{{\mathbf{X}}^{T}}\bm{\kappa }{{\mathbf{x}}_{i}}.
\end{equation}

Then, we can get the representation of nonlocal linear regression model with ${{h}_{i,j}}=\omega _{i}^{j}$:
\begin{equation}
\mathbf{x}=\mathbf{Hx}+\mathbf{e}.
\end{equation}
Where $\mathbf{e}$ is the modeling error. We incorporate the above model into image interpolation model, the Eq. (2) can be rewritten as:
\begin{equation}
\underset{\bm{\alpha }}{\mathop{\min }}\,\left\{ \left\| \mathbf{y}-\mathbf{DH\Phi}\bm{\alpha } \right\|_{2}^{2}+\lambda R(\bm{\alpha }) \right\}\begin{matrix}
   {} & s.t. & \mathbf{y}=\mathbf{D\Phi}\bm{\alpha }  \\
\end{matrix}.
\end{equation}

\subsection{Regularization terms}
To solve the model formulated in Eq. (9), two widely used priors including image sparsity and non-local self-similarity are used as the regularization terms. The sparsity of coding coefficient ${{\bm{\alpha }}_{i}}$ can be characterized by ${{\left\| {{\bm{\alpha }}_{i}} \right\|}_{0}}$ \cite{13, 14, 27, 28}. Non-local self-similarity prior refers to the fact that a local image patch often has many non-local similar patches to it across the image for natural images. These non-local similar patches may be spatially either close to or far from this patch. This prior has been successfully adopted in various applications of image restoration \cite{18,19,21,29,30,31}.

For each patch ${{\mathbf{x}}_{i}}$ of size $p\times p$, we can get its first $t$ most similar patches by calculating the Euclidean distance between them. After we get the similar patches, denoted by $\{\mathbf{x}_{i}^{1},\mathbf{x}_{i}^{2},\ldots ,\mathbf{x}_{i}^{t}\}$, these similar patches can be used to estimate ${{\mathbf{x}}_{i}}$ by weighted average: ${{\widehat{\mathbf{x}}}_{i}}=\sum\nolimits_{k=1}^{t}{{{a}_{k}}\mathbf{x}_{i}^{k}}$. Term ${{a}_{k}}$ is the weighted coefficient which is inversely proportional to the distance: ${{a}_{k}}={\exp ({-\left\| {{\mathbf{x}}_{i}}-\mathbf{x}_{i}^{k} \right\|_{2}^{2}}/{{{h}_{1}}}\;)}/{{{h}_{2}}}\;$, where ${{h}_{1}}$ is a preset scalar and ${{h}_{2}}$ is a normalization factor. For a given dictionary ${{\mathbf{\Phi }}_{i}}$, we can get the sparse representations of patch ${{\mathbf{x}}_{i}}$ and prediction ${{\widehat{\mathbf{x}}}_{i}}$, i.e. ${{\mathbf{x}}_{i}}={{\mathbf{\Phi }}_{i}}{{\bm{\alpha }}_{i}}$ and ${{\widehat{\mathbf{x}}}_{i}}={{\mathbf{\Phi }}_{i}}{{\bm{\beta }}_{i}}$. This two sparse representations should be as close as possible. In other words, the difference between ${{\bm{\alpha }}_{i}}$ and ${{\bm{\beta }}_{i}}$ should be as smaller as possible. Then, image sparsity prior $\sum\nolimits_{i}{{{\left\| {{\bm{\alpha }}_{i}} \right\|}_{0}}}$ and nonlocal self-similarity prior $\sum\nolimits_{i}{\left\| {{\bm{\alpha }}_{i}}-{{\bm{\beta }}_{i}} \right\|_{2}^{2}}$ are integrated into the above model. The interpolation model can be described as following:
\begin{equation}
\begin{array}{l}
\mathop {\min }\limits_{\bm{\alpha }} \left\{ {\left\| {{\bf{y}} - {\bf{DH\Phi}\bm{ \alpha }}} \right\|_2^2 + \lambda \sum\limits_{i = 1}^K {{{\left\| {{{\bm{\alpha }}_i}} \right\|}_0}} } \right.\\
\left. { + \eta \sum\limits_{i = 1}^K {\left\| {{{\bm{\alpha }}_i} - {{\bm{\beta }}_i}} \right\|_2^2} } \right\}\begin{array}{*{20}{c}}
{}&{s.t.}&{{\bf{y}} = {\bf{D\Phi}\bm{ \alpha }}}.
\end{array}
\end{array}
\end{equation}
Where $K$ is the patch number partitioned from the image ${\mathbf{x}}$ and $\eta $ is the regularization parameter. Since the reweighted ${{\ell }_{\text{p}}}$ norm can enhance the sparsity and get a better solution \cite{32}, the reweighted strategy is integrated into the above model:
\begin{equation}
\begin{array}{c}
\mathop {\min }\limits_{\bm{\alpha }} \left\{ {\left\| {{\bf{y}} - {\bf{DH\Phi}\bm{ \alpha }}} \right\|_2^2 + \lambda \sum\limits_{i = 1}^K {{{\left\| {{{\bm{\alpha }}_i}} \right\|}_0}} } \right.\\
\left. { + \sum\limits_{i = 1}^K {\sum\limits_{j = 1}^{{p^2}} {{\eta _{i,j}}{{({\alpha _{i,j}} - {\beta _{i,j}})}^2}} } } \right\}\begin{array}{*{20}{c}}
{}&{s.t.}&{{\bf{y}} = {\bf{D\Phi}\bm{ \alpha }}}.
\end{array}
\end{array}
\end{equation}
Where ${{\alpha }_{i,j}},{{\beta }_{i,j}}$ are the ${{j}^{th}}$ element of vectors ${{\bm{\alpha }}_{i}}$ and ${{\bm{\beta }}_{i}}$ respectively. 

\subsection{Adaptive dictionary selection}
The selection of dictionary plays an important role in the reconstruction of a signal. Lots of researches have shown to learn a universal and over-complete dictionary \cite{13,14,15,16}. However, this dictionary is not effective and optimal because many atoms are irrelevant to a given local patch. Adaptive sub-dictionaries are adopted in \cite{17,18,19,20}. There are two methods to train sub-dictionaries: pre-trained and online trained. Abundant high-quality images are needed for pre-trained dictionaries. However, pre-trained dictionaries are not always valid if they are irrelevant to the content of a given patch. Online trained sub-dictionaries in \cite{18} refers to that image patches are clustered and a PCA sub-dictionary is trained for each class. However, the class number and sample number of each class are difficulty to set. Inadequate clustering will reduce the accuracy of sparse representation, leading to the inaccuracy of reconstructed signals.

In this paper, we propose a new sub-dictionary trained method. As described in non-local self-similarity priors, for each patch ${{\mathbf{x}}_{i}}$, we can get its similarity patches in a large enough local window. A patch $\mathbf{x}_{i}^{k}$ is selected as a similar patch to ${{\mathbf{x}}_{i}}$ if the Euclidean distance between them is not greater than the preset threshold. All similar patches are sorted by the Euclidean distance in ascending order. We choose the first $n$ patches as training samples to train a PCA sub-dictionary. The term $n$ should be large enough to guarantee a reasonable training.

These training samples are donated as $\mathbf{x}_{i}^{0},\mathbf{x}_{i}^{1},\mathbf{x}_{i}^{2},\ldots \mathbf{x}_{i}^{n-1}$. They are grouped as a matrix ${{\mathbf{X}}_{i}}=[\mathbf{x}_{i}^{0},\mathbf{x}_{i}^{1},\mathbf{x}_{i}^{2},\ldots \mathbf{x}_{i}^{n-1}]$. Next we calculate the covariance matrix of ${{\mathbf{X}}_{i}}$, denoted by ${{\mathbf{C}}_{{{\mathbf{X}}_{i}}}}$. Finally we can get the adaptive PCA sub-dictionary ${{\mathbf{\Phi }}_{i}}$ of the patch ${{\mathbf{x}}_{i}}$ by calculating the eigenvalue decomposition of ${{\mathbf{C}}_{{{\mathbf{X}}_{i}}}}$:
\begin{equation}
{{\mathbf{\Phi }}_{i}}={{\mathbf{Q}}^{T}}.
\end{equation}
Where $\mathbf{Q}$ is the eigenvector matrix and ${{\mathbf{C}}_{{{\mathbf{X}}_{i}}}}=\mathbf{Q\Sigma }{{\mathbf{Q}}^{T}}$. In fact, these adaptive PCA sub-dictionaries form an over-complete dictionary $\mathbf{\Phi }=[{{\mathbf{\Phi }}_{1}},{{\mathbf{\Phi }}_{2}},{{\mathbf{\Phi }}_{3}},\ldots ,{{\mathbf{\Phi }}_{K}}]$ for image $\mathbf{x}$. For a given patch ${{\mathbf{x}}_{i}}$, the adaptive dictionary ${{\mathbf{\Phi }}_{i}}$ is selected to code ${{\mathbf{x}}_{i}}$. This makes the sparse coding coefficient of ${{\mathbf{x}}_{i}}$ over the rest sub-dictionaries be zero, leading to a very sparse representation of ${{\mathbf{x}}_{i}}$. So our method will ensure the sparsity of coding coefficient and the sparse regularization term can be removed. Thus, over adaptive sub-dictionaries, the image interpolation model can be described as:
\begin{equation}
\begin{array}{l}
\widehat {\bm{\alpha }} = \arg \mathop {\min }\limits_{\bm{\alpha }} \left\{ {\left\| {{\bf{y}} - {\bf{DHx}}} \right\|_2^2 + \gamma \sum\limits_{i = 1}^K {\left\| {{{\bf{R}}_i}{\bf{x}} - {{\bf{\Phi }}_i}{{\bm{\alpha }}_i}} \right\|_2^2} } \right.\\
\left. { + \sum\limits_{i = 1}^K {\sum\limits_{j = 1}^{{p^2}} {{\eta _{i,j}}{{({\alpha _{i,j}} - {\beta _{i,j}})}^2}} } } \right\}\begin{array}{*{20}{c}}
{}&{s.t.}&{{\bf{y}} = {\bf{Dx}}}.
\end{array}
\end{array}
\end{equation}

\subsection{Weighted encoding}
Here we discuss the ${{\ell }_{2}}$ norm data fidelity term $\left\| \mathbf{y}-\mathbf{DHx} \right\|_{2}^{2}$ in order to get a better solution of Eq. (13). Since the ${{\ell }_{2}}$ norm is optimal for Gaussian distribution, a non-Gaussian distribution of fitting residuals $(\mathbf{y}-\mathbf{DH}{{\mathbf{x}}^{(s)}})$ will not produce an optimal solution, where ${{\mathbf{x}}^{(s)}}$ is the estimation value of $\mathbf{x}$ in the ${{s}^{th}}$ iteration and it can be obtained by ${{\mathbf{x}}^{(s)}}=\mathbf{\Phi }\widehat{\bm{\alpha }}$. Here we use an example to investigate the distribution of fitting residuals on image \emph{House}.
\begin{figure}[h]
\centering
\includegraphics[width=\linewidth]{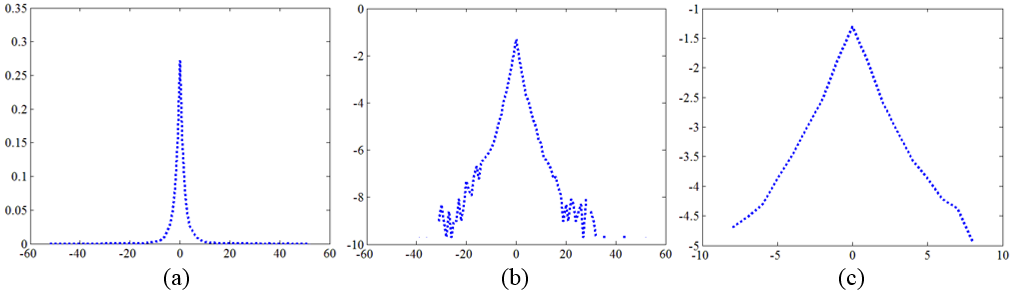}
\caption{The distribution of fitting residual in (a) linear and (b) log domain, (c) weighted fitting residual in log domain.}
\end{figure}

Figure 1(a) and (b) show the distributions of fitting residuals in linear and log domain respectively. As shown in Fig. 1 (b), one can see that the distribution of fitting residuals doesn't follow regular Gaussian distribution and it has a tail. Thus, using the ${{\ell }_{2}}$ norm to characterize the data fidelity term is not optimal. In order to weaken the effect of tail, we can modify the fitting residual by assigning a proper weight so that its distribution can be more like Gaussian distribution. Then the ${{\ell }_{2}}$ norm can be still used to characterize the data fidelity term. The fitting residuals can be used to guide the setting of weights because tail exists at high residuals. Thus, the weights should be inversely proportional to the fitting residuals. Hereby we give a simple and effective choice of weights $\mathbf{w}$ as:
\begin{equation}
\mathbf{w}=\exp (-{{c}_{1}}(\mathbf{y}-\mathbf{DH}{{\mathbf{x}}^{(s)}})\odot (\mathbf{y}-\mathbf{DH}{{\mathbf{x}}^{(s)}})).
\end{equation}
Where ${{c}_{1}}$ is positive controlling constant and $\odot $ represents component-wise multiplication. Figure 1 (c) shows the distribution of weighted residuals in log domain. We can see that the distribution of weighted residuals is more like Gaussian distribution, which will guarantee that the ${{\ell }_{2}}$ norm can be still used in the data fidelity term. Then the model formulated in Eq. (13) can be modified as:
\begin{equation}
\begin{array}{l}
\mathop {\min }\limits_{\bm{\alpha }} \left\{ {\left\| {{{\bf{W}}^{1/2}}({\bf{y}} - {\bf{DHx}})} \right\|_2^2 + \gamma \sum\limits_{i = 1}^K {\left\| {{{\bf{R}}_i}{\bf{x}} - {{\bf{\Phi }}_i}{{\bm{\alpha }}_i}} \right\|_2^2} } \right.\\
\left. { + \sum\limits_{i = 1}^K {\sum\limits_{j = 1}^{{p^2}} {{\eta _{i,j}}{{({\alpha _{i,j}} - {\beta _{i,j}})}^2}} } } \right\}\begin{array}{*{20}{c}}
{}&{s.t.}&{{\bf{y}} = {\bf{Dx}}}.
\end{array}
\end{array}
\end{equation}
Where $\mathbf{W}=diag(\mathbf{w})$.

\subsection{Interpolation Algorithm}
Given a current estimate of image $\mathbf{x}$, the weighted encoding matrix $\mathbf{W}$ and the sparse representation ${{\bm{\alpha }}_{i}}$ can be updated in the next iteration. Then, the updated $\mathbf{W},{{\mathbf{\alpha }}_{i}}$ and dictionary $\mathbf{\Phi }$ are in turn used to update the estimate of image $\mathbf{x}$. Such an iteration is stopped until a stopping rule is met.

For a given $\mathbf{x}$, the Eq. (15) can be rewritten by only retaining the components related to ${{\bm{\alpha }}_{i}}$ as:
\begin{equation}
\begin{aligned}
  {{\widehat{\bm{\alpha }}}_{i}}&=\arg \underset{{{\bm{\alpha }}_{i}}}{\mathop{\min }}\,J({{\bm{\alpha }}_{i}}) \\ 
 & =\gamma \left\| {{\mathbf{R}}_{i}}\mathbf{x}-{{\mathbf{\Phi }}_{i}}{{\bm{\alpha }}_{i}} \right\|_{2}^{2}+\sum\limits_{j=1}^{{{p}^{2}}}{{{\eta }_{i,j}}{{({{\alpha }_{i,j}}-{{\beta }_{i,j}})}^{2}}}  
\end{aligned}
\end{equation}
This is a quadratic problem and we can get a close-form solution by setting ${\partial J}/{\partial }\;{{\bm{\alpha }}_{i}}=0$.
\begin{equation}
\alpha _{i,j}^{(s+1)}=\frac{{{(\mathbf{\Phi }_{i}^{T}{{\mathbf{R}}_{i}}{{\mathbf{x}}^{(s)}})}_{j}}+{{{\eta }_{i,j}}{{\beta }_{i,j}}}/{\gamma }\;}{1+{{{\eta }_{i,j}}}/{\gamma }\;}.
\end{equation}

After obtaining sparse representations, the image $\mathbf{x}$ can be optimized by minimizing the following function:
\begin{equation}
\begin{array}{l}
\widehat {\bf{x}} = \arg \mathop {\min }\limits_{\bf{x}} \left\{ {\left\| {{{\bf{W}}^{1/2}}({\bf{y}} - {\bf{DHx}})} \right\|_2^2} \right.\\
\left. { + \gamma \sum\limits_{i = 1}^K {\left\| {{{\bf{R}}_i}{\bf{x}} - {{\bf{\Phi }}_i}{{\bf{\alpha }}_i}} \right\|_2^2} } \right\}\begin{array}{*{20}{c}}
{}&{s.t.}&{{\bf{y}} = {\bf{Dx}}}.
\end{array}
\end{array}
\end{equation}
To solve the above optimization problem, the Augmented Lagrange Multiplier \cite{33,34} algorithm can be used. The augmented Lagrangian function can be defined as:
\begin{equation}
\begin{array}{l}
{\cal L}({\bf{x}},{\bf{f}},\mu ) = \left\| {{{\bf{W}}^{1/2}}({\bf{y}} - {\bf{DHx}})} \right\|_2^2 + \gamma \sum\limits_{i = 1}^K {\left\| {{{\bf{R}}_i}{\bf{x}} - {{\bf{\Phi }}_i}{{\bm{\alpha }}_i}} \right\|_2^2} \\
 + \left\langle {{\bf{f}},{\bf{y}} - {\bf{Dx}}} \right\rangle  + \mu \left\| {{\bf{y}} - {\bf{Dx}}} \right\|_2^2
\end{array}
\end{equation}
We can get the estimate ${{\mathbf{x}}^{(s+1)}}$ of image ${\mathbf{x}}$ in the ${{s}^{th}}$ iteration by setting ${\partial \mathcal{L}}/{\partial \mathbf{x}}\;=0$.
\begin{equation}
\begin{array}{l}
{{\bf{x}}^{(s + 1)}} = {\left[ {{{({\bf{DH}})}^T}{{\bf{W}}^{(s)}}({\bf{DH}}) + \gamma \sum\limits_{i = 1}^K {{\bf{R}}_i^T{{\bf{R}}_i}}  + {\mu ^{(s)}}{{\bf{D}}^T}{\bf{D}}} \right]^{ - 1}}\\
\left[ {{{({\bf{DH}})}^T}{{\bf{W}}^{(s)}}{\bf{y}} + \gamma \sum\limits_{i = 1}^K {{\bf{R}}_i^T({{\bf{\Phi }}_i}{{\bm{\alpha }}_i})}  + {\mu ^{(s)}}{{\bf{D}}^T}{\bf{y}} + {{{{\bf{D}}^T}{{\bf{f}}^{(s)}}} \mathord{\left/
 {\vphantom {{{{\bf{D}}^T}{{\bf{f}}^{(s)}}} 2}} \right.
 \kern-\nulldelimiterspace} 2}} \right]
\end{array}
\end{equation}
Where ${{\mathbf{f}}^{(s+1)}}={{\mathbf{f}}^{(s)}}+{{\mu }^{(s)}}(\mathbf{y}-\mathbf{D}{{\mathbf{x}}^{(s+1)}})$ and ${{\mu }^{(s+1)}}=t\cdot {{\mu }^{(s)}}$ with a constant $t>1$. Here, the overall interpolation algorithm is summarized in \textbf{Algorithm 1}.
\begin{algorithm}[h]
\caption{Weighted Encoding Based Image Interpolation With Nonlocal Linear Regression Model} 
{\bf Input:}
Low-resolution image $\mathbf{y}$.\\
 {\bf Initialization:}
Initialize iterator $s=0$, maximum iteration number $T$ and set:\\
\hspace*{0.02in} \  the initial high-resolution image $\mathbf{x}$ using the Bicubic  inter-\\
\hspace*{0.02in} \  polation method.\\
\hspace*{0.02in} \  the related parameters: $\eta =1.2,\gamma =0.1,\mu =0.68,t=1.1,$\\
\hspace*{0.02in} \  $m=15,t=23$, image block size: $5\times 5$ and ${{\mathbf{f}}^{(0)}}=\mathbf{0}$.\\
\hspace*{0.02in} \  PCA sub-dictionaries learned according to the Section II.C\\
{\bf Main Iteration:}
Increment $s$ by 1 and perform the following steps:\\
1. Calculate sparse representation $\{{{\bm{\alpha }}_{i}}\}$ by Eq. (17).\\
2. Calculate image ${{\mathbf{x}}^{(s+1)}}$ by Eq. (20).\\
3. Update ${{\mathbf{f}}^{(s+1)}}$ and ${{\mu }^{(s+1)}}$, update weighted matrix ${{\mathbf{W}}^{(s+1)}}$ by Eq. (14).\\
4. Update parameters:  ${{\eta }_{i,j}}={{{{{k}_{2}}}/{((\alpha _{i,j}^{(s+1)}-{{\beta }_{i,j}})}\;}^{2}}+\varepsilon )$ according to the weight assigned strategy described in \cite{31}. \\
5. Update the adaptive dictionaries $\mathbf{\Phi }=[{{\mathbf{\Phi }}_{1}},{{\mathbf{\Phi }}_{2}},{{\mathbf{\Phi }}_{3}},\ldots ,{{\mathbf{\Phi }}_{K}}]$ by Eq. (12).\\
6. Stopping rule: if $s>T$, stop. Otherwise, do another iteration.\\
{\bf Output:}
High-resolution image $\mathbf{x}$.
\end{algorithm}

\section{Experimental results}
In all experiments, the size of an image patch is set to $5\times 5$ pixels and the controlling constant ${{c}_{1}}$ is set to $0.006$. Ten benchmark images shown in Fig. 2 are used to evaluate the interpolation performance. These benchmark images are down-sampled firstly. Then low-resolution images are interpolated to generate high-resolution interpolated images that are compared with benchmark images in terms of quantitative measures and visual quality. The PSNR, SSIM \cite{35} and FSIM \cite{36} are used as quantitative evaluation criterions. We compare our proposed method with several state-of-the art image interpolation methods including: NEDI \cite{4}, DFDF \cite{5}, and sparse representation based methods including  SME \cite{7}, SCSR \cite{16} and NARM \cite{18}. All the source codes are downloaded from the homepage of the corresponding authors. 
 
\begin{figure*}[!ht]
\centering
\includegraphics[width=0.8\linewidth]{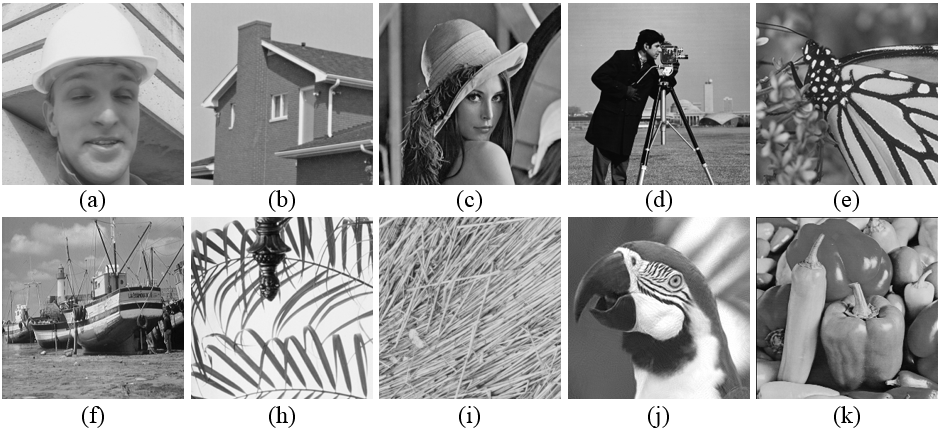}
\caption{The ten benchmark images. From left to right and top to bottom: \emph{Foreman, House, Lena, Cameraman, Monarch, Boat, Leaves, Straw, Parrot and Peppers}.}
\end{figure*}

\begin{figure*}[!ht]
\centering
\includegraphics[width=0.8\linewidth]{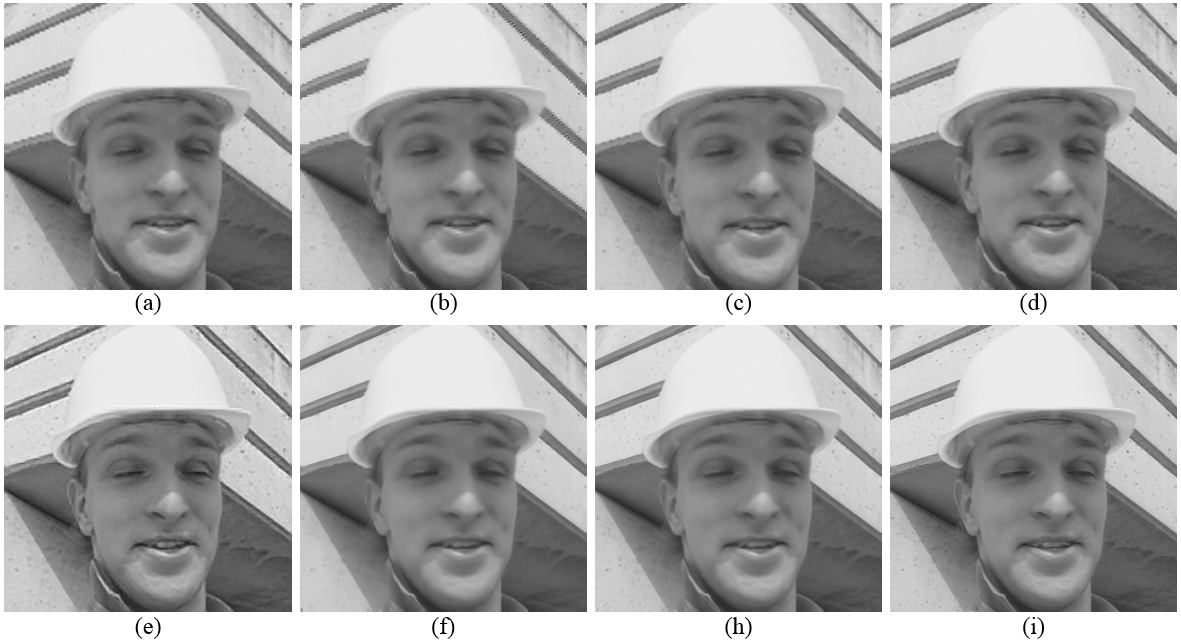}
\caption{Interpolated \emph{Foreman} images with upscaling factor of 2. (a) Bicubic (b) NEDI (c) DFDF (d) SME (e) SCSR (f) NARM (h) Ours (I) True image.}
\end{figure*}

\begin{figure*}[!ht]
\centering
\includegraphics[width=0.6\linewidth]{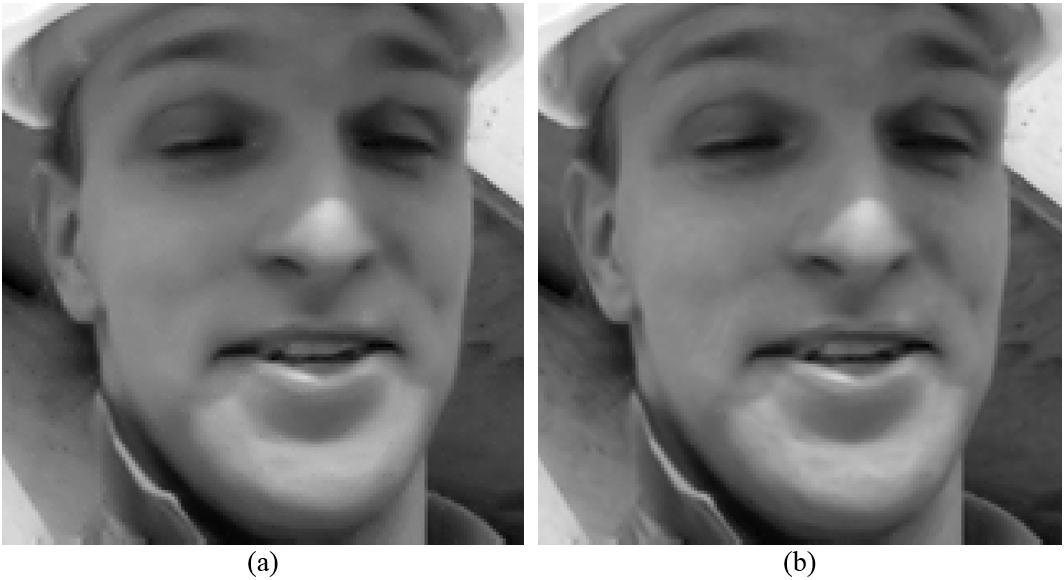}
\caption{Comparison of interpolated results. (a) NARM (b)Ours.}
\end{figure*}

\begin{figure*}[!ht]
\centering
\includegraphics[width=0.8\linewidth]{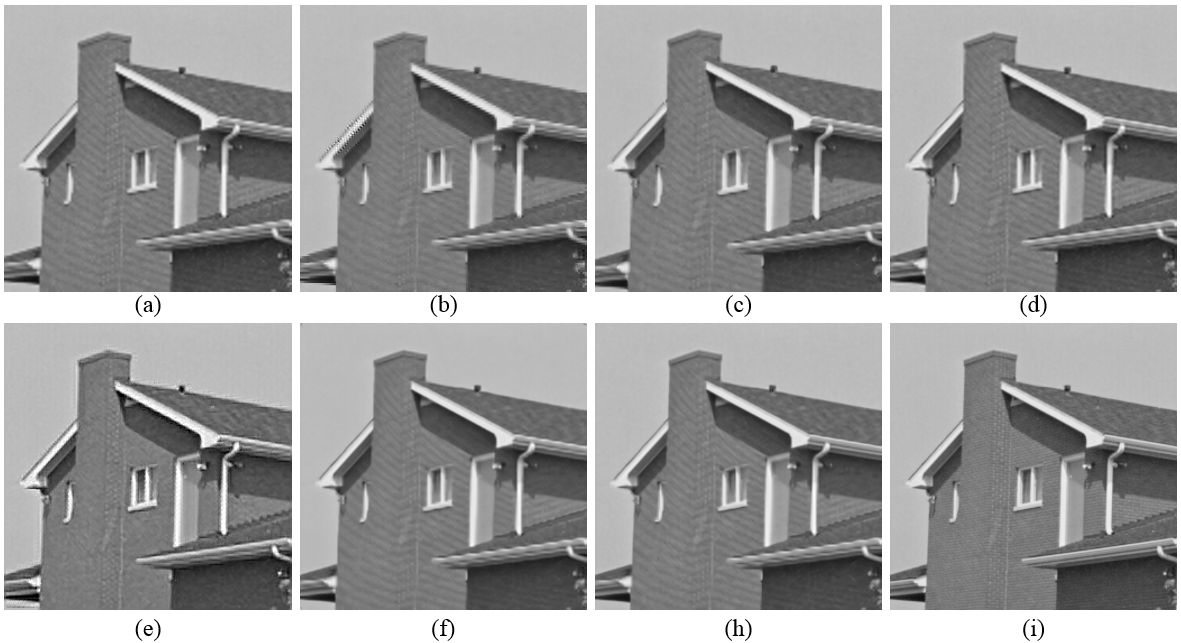}
\caption{Interpolated \emph{House} images with upscaling factor of 2. (a) Bicubic (b) NEDI (c) DFDF (d) SME (e) SCSR (f) NARM (h) Ours (I) True image.}
\end{figure*}

\begin{figure*}[!ht]
\centering
\includegraphics[width=0.8\linewidth]{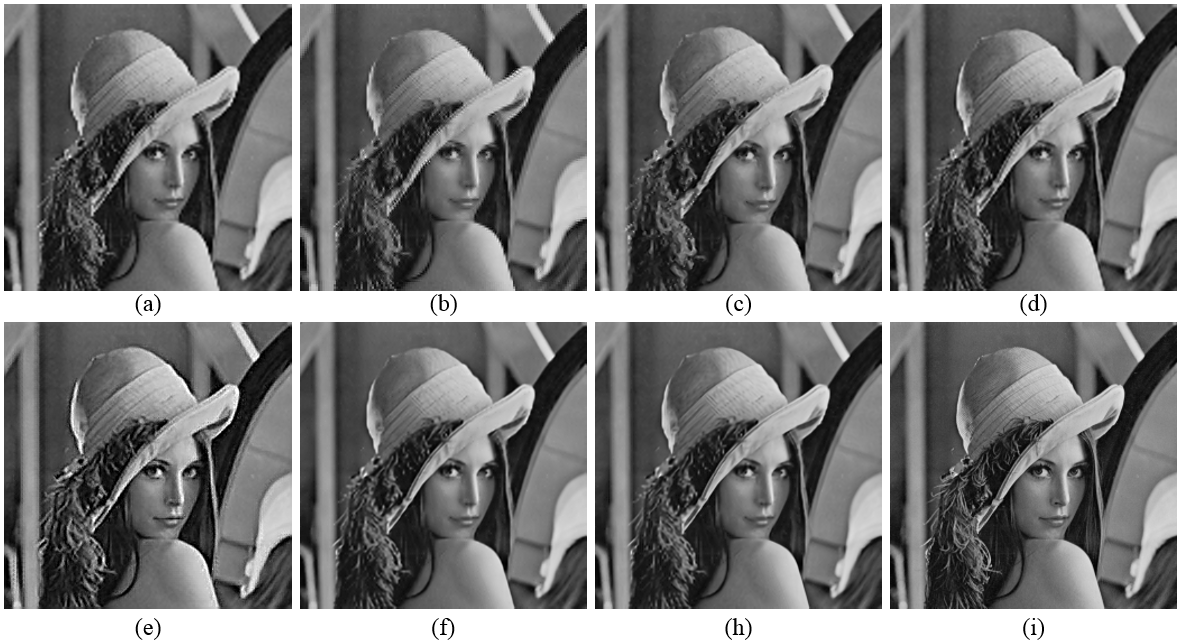}
\caption{Interpolated \emph{Lena} images with upscaling factor of 2. (a) Bicubic (b) NEDI (c) DFDF (d) SME (e) SCSR (f) NARM (h) Ours (I) True image.}
\end{figure*}

\begin{table*}[!t]
\centering
\caption{The PSNR, SSIM, FSIM results on ten test images by different methods with upscaling factor of 2.}
\begin{tabular}{c|c|c|c|c|c|c|c}
\hline
\textbf{Image} & Bicubic & NEDI & DFDF & SCSR & SME & NARM & Ours\\
\hline
\multirow{3}*{Foreman}   &35.2710	&33.1125	&36.6937	&32.2409	&36.6597	&38.3884	&\textbf{38.5234}\\
				       &0.94661	&0.93475	&0.95392	&0.91756	&0.95296	&0.95675	&\textbf{0.95709}\\
				       &0.96411	&0.95723	&0.97099	&0.93464	&0.97089	&0.97472	&\textbf{0.97681}\\
\hline
\multirow{3}*{House}        &32.2486	&30.8612	&32.6265	&29.1428	&33.1718	&33.5411	&\textbf{33.6836}\\
				       &0.88056	&0.87772	&0.87941	&0.86377	&0.88415	&\textbf{0.88808}	&0.88725\\
				       &0.93925	&0.93168	&0.94745	&0.89903	&0.94978	&0.95335	&\textbf{0.95718}\\
\hline
\multirow{3}*{Lena}        &29.3366	&28.3854	&29.5212	&26.8893	&30.0699	&30.5373	&\textbf{30.6419}\\
				      &0.90076	&0.88702	&0.89846	&0.86495	&0.90840	&\textbf{0.91514}	&0.91422\\
				      &0.95155	&0.94223	&0.95330	&0.91567	&0.95674	&0.96089	&\textbf{0.96120}\\
\hline
\multirow{3}*{Cameraman}        &25.5058	&25.0664	&25.7028	&24.9155	&\textbf{26.2285}	&26.0238	&26.1372\\
				      &0.85946	&0.84911	&0.86693	&0.84639	&0.86588	&\textbf{0.87286}	&0.87091\\
				      &0.90165	&0.89277	&0.91380	&0.87300	&0.90808	&0.91592	&\textbf{0.91790}\\
\hline
\multirow{3}*{Monarch}       &28.4124	&26.8815	 &29.0283	&24.8682	&29.3786	&30.6136	&\textbf{30.7062}\\
&0.93332 &0.91324	&0.93966	&0.86188	&0.94480	&0.95443	&\textbf{0.95585}\\
&0.93404	&0.92397	&0.95093	&0.85387	&0.94267	&0.96114	&\textbf{0.96324}\\
\hline
\multirow{3}*{Boat}       &23.8265	&23.8807	&23.9313	&23.4562	&24.1617	&24.0235	&\textbf{24.1747}\\
&0.75894	&0.75954	&0.75768	&0.76690	&0.76966	&\textbf{0.77340}	&0.77032\\
&0.89220	&0.88846	&0.89414	&0.86398	&0.89581	&0.89936	&\textbf{0.90055}\\
\hline
\multirow{3}*{Leaves}       &26.6446	&23.6711	&27.0701	&22.6739	&27.8381	&29.4344	&\textbf{29.7586}\\
&0.93511	&0.88891	&0.94297	&0.85739	&0.94891	&0.96559	&\textbf{0.96691}\\
&0.92537	&0.89790	&0.94687	&0.84674	&0.94048	&0.96437	&\textbf{0.96813}\\
\hline
\multirow{3}*{Straw}       &24.9503	&22.8035	&24.5504	&21.9469	&26.2127	&\textbf{27.1254}	&27.1181\\
&0.83166	&0.76751	&0.81380	&0.80407	&0.87882	&\textbf{0.90318}	&0.90256\\
&0.91706	&0.88674	&0.90933	&0.87029	&0.93218	&0.94247	&\textbf{0.94255}\\
\hline
\multirow{3}*{Parrot}       &29.9697	&28.8591	&29.8211	&27.5569	&30.5317	&\textbf{31.3254}	&31.1769\\
&0.85128	&0.85321	&0.85775	&0.74726	&0.85618	&\textbf{0.87527}	&0.86273\\
&0.94709	&0.94436	&0.94744	&0.88352	&0.95037	&\textbf{0.95445}	&0.95308\\
\hline
\multirow{3}*{Peppers}       &28.3370 &24.0673	&\textbf{30.4757}	&22.5981	&29.2094	&29.7449	&30.1449\\
&0.91131	&0.90219	&0.91690	&0.85086	&0.91690	&\textbf{0.92685}	&0.92602\\
&0.94120	&0.92159	&0.95447	&0.93708	&0.94644	&0.95440	&\textbf{0.95586}\\
\hline
\multirow{3}*{\textbf{Average}}       &28.4503	&26.7589	&28.9421	&25.9183	&29.3462	&30.0758	&\textbf{30.2066}\\
&0.88090	&0.86332	&0.88275	&0.84983	&0.89267	&\textbf{0.90315}	&0.90138\\
&0.93135	&0.91869	&0.93887	&0.88945	&0.93934	&0.94811	&\textbf{0.94964}\\
\hline
\end{tabular}
\end{table*}

\begin{figure*}[!ht]
\centering
\includegraphics[width=0.8\linewidth]{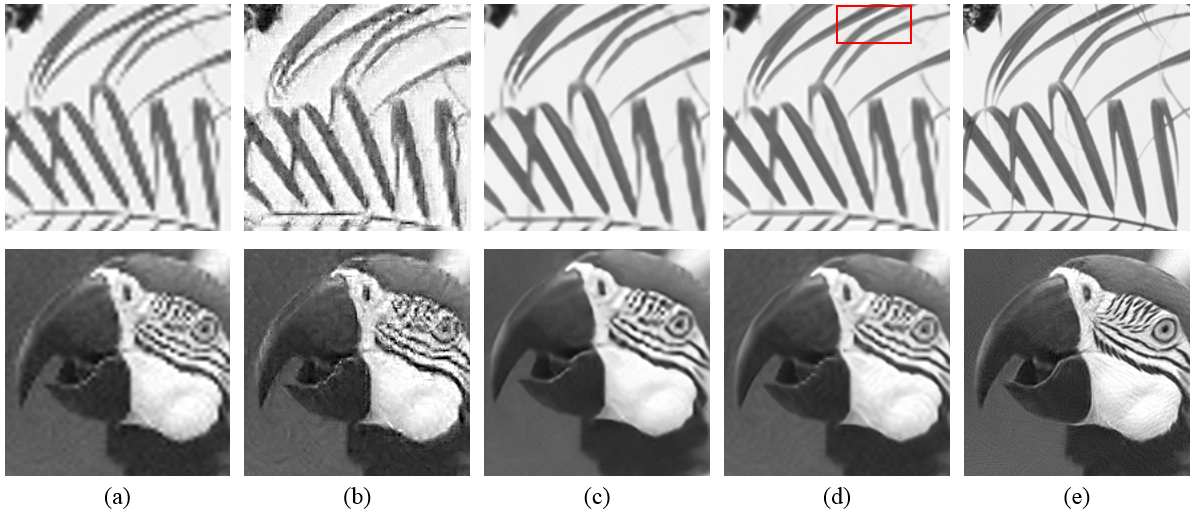}
\caption{Interpolated images with upscaling factor of 3. (a) Bicubic (b) SCSR (c) NARM (d) Ours (e) True image.}
\end{figure*}

\begin{table*}[!t]
\centering
\caption{The PSNR, SSIM, FSIM results on ten test images by different methods with upscaling factor of 3.}
\begin{tabular}{c|c|c|c|c|c|c|c|c|c|c|c}
\hline
\textbf{Image} & Foreman & House & Lena & Cameraman & Monarch & Boat & Leaves & Straw & Parrot & Peppers & \textbf{Average}\\
\hline
\multirow{3}*{Bicubic} &31.8636	&28.7770	&25.9538	&22.5412	&24.2393	&21.6294	&21.7580	&21.0395	&26.3175	&25.2435	&24.9363\\
&0.90333	&0.82423	&0.81964	&0.76651	&0.84514	&0.62445	&0.81254	&0.60056	&0.79316	&0.84831	&0.78379\\
&0.92655	&0.87769	&0.90303	&0.81881	&0.85563	&0.81557	&0.81382	&0.81657	&0.91292	&0.88894	&0.86295\\
\hline
\multirow{3}*{SCSR} &29.2492	&26.6055	&23.7016	&21.4478	&21.8788	&20.4430	&19.1820	&18.8389	&24.3585	&21.0116	&22.6717\\
&0.85242	&0.76605	&0.72696	&0.70668	&0.74839	&0.57409	&0.70170	&0.50466	&0.73789	&0.73761	&0.70565\\
&0.90159	&0.84642	&0.85850	&0.78663	&0.78499	&0.79184	&0.75651	&0.79136	&0.88878	&0.81078	&0.82174\\
\hline
\multirow{3}*{NARM} &29.0512	&28.4730	&26.2937	&22.3291	&25.2467	&21.4987	&22.2139	&20.9125	&26.5037	&24.5912	&24.7114\\
&0.91237	&\textbf{0.83704}	&\textbf{0.84115}	&0.77954	&\textbf{0.88830}	&\textbf{0.64287}	&0.87243	&\textbf{0.64142}	&\textbf{0.82828}	&0.86744	&\textbf{0.81108}\\
&0.92570	&0.86933	&0.90222	&0.80844	&0.89907	&0.80702	&0.88036	&\textbf{0.82922}	&0.90833	&0.90075	&0.87304\\
\hline
\multirow{3}*{Ours} &\textbf{34.7259}	&\textbf{29.7260}	&\textbf{26.6696}	&\textbf{22.7528}	&\textbf{25.8284}	&\textbf{21.8178}	&\textbf{23.3044}	&\textbf{21.3857}	&\textbf{27.0098}	&\textbf{27.8500}	&\textbf{26.1070}\\
&\textbf{0.92964}	&0.83699	&0.83694	&\textbf{0.78698}	&0.88628	&0.63532	&\textbf{0.87827}	&0.63047	&0.81817	&\textbf{0.87067}	&0.81097\\
&\textbf{0.95048}	&\textbf{0.89986}	&\textbf{0.91145}	&\textbf{0.84395}	&\textbf{0.90540}	&\textbf{0.81997}	&\textbf{0.88954}	&0.81095	&\textbf{0.92058}	&\textbf{0.91832}	&\textbf{0.88705}\\
\hline
\end{tabular}
\end{table*}

Interpolated \emph{foreman} images with the upscaling factor of 2 are shown in Fig. 3. The Bicubic method generates severe artifacts along edges. Those edge based methods including NEDI and DFDF produce better interpolated results and weaken artifacts, but these methods can't remove artifacts effectively because it is difficulty to estimate the edge direction from low-resolution images. The SME and SCSR methods work much better in preserving edges than edge based methods. However, these methods still can't generate sharp edges and the SCSR method produces phantom artifacts along edge.  The NARM produces a better interpolated results than above methods and it preserves image edges. However, the learned dictionaries are based on clustering in this method and inappropriate cluster generates spackle noise as shown in Fig. 4. In our method, similar patches are grouped to train adaptive dictionaries instead of clustering, producing a more sparse and accurate representation. This guarantees that our method can produce much better results as shown in Fig. 3(h). For better view spackle noise, the cropped portions of \emph{foreman} image are shown in Fig. 4. One can see that there are severe spackle noise in teeth and ears. Our proposed method produce more clean results while preserving edges. Interpolated \emph{house} and \emph{lena} images with upscaling factor of 2 are shown in Fig. 5 and Fig. 6 respectively. We can see that our proposed method produce much better results than other methods. Our method can reconstruct more clean images while preserving edges such as eaves in Fig. 5 and brims of a hat in Fig. 6.

More interpolated results obtained with upscaling factor of 2 by different methods are list in Table I. The PSNR, SSIM and FSIM results are included in Table I. For each image, from top to bottom are PSNR, SSIM and FSIM values. The bold number is the best result in each row. We can see that our method achieves higher PSNR, SSIM and FSIM for most benchmark images, which demonstrates that our method is superior to other methods in terms of quantitative measures. 

We also conduct experiments with upscaling factor of $3$. Since NEDI, DFDF, SME methods are designed for upscaling factor of ${{2}^{n}}$, we just compare our proposed method with Bicubic, SCSR and NARM algorithms. The cropped portions of interpolated images with upscaling factor of $3$ are shown in Fig. 7 and quantitative
measures are list in Table II. For \emph{Leaves} image, we can see that the Bicubic method generates severe artifacts. The SCSR method can weaken artifacts, but it can't generate sharp edges while reconstructing severe phantom. The NARM method is better than the first two methods. However, our proposed method preserves edges better such as in the red rectangle region. For \emph{Parrot} image, the NARM produce much smooth results than ours, but it has lower PSNR and FSIM values.  According to results list in Table II, one can see that our proposed method achieves higher PSNR, SSIM and FSIM for most benchmark images, which demonstrates that our method is superior to other methods in terms of quantitative measures.

\section{Conclusion}
In this paper, we propose a novel image interpolation model based on sparse representation. In order to improve interpolation model, we incorporate nonlocal linear regression into it and adopt nonlocal self-similar prior as the regularization term. Besides, we propose a new approach to learn adaptive sub-dictionaries, which ensure that coding coefficients more sparse and accurate. Moreover, we introduce weighted encoding into data fidelity to suppress tailing of fitting residuals. Abundant benchmark images are used to evaluate the interpolation performance. Experimental results demonstrate that our proposed method outperforms several state-of-the-art methods in terms of quantitative measures and visual quality.

\ifCLASSOPTIONcaptionsoff
  \newpage
\fi



\bibliographystyle{IEEEtran}
\bibliography{IEEEfull,mybibfile}
%



%

\begin{IEEEbiography}[{\includegraphics[width=1in,height=1.25in]{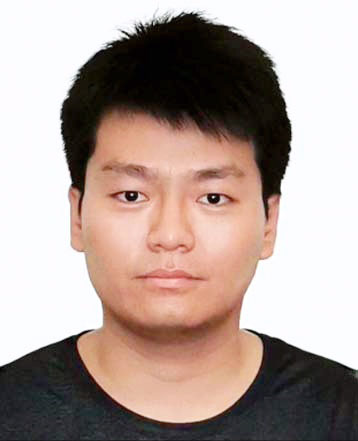}}]{Junchao Zhang}
received the B.S. degree in mechanical engineering and automation from HoHai University in 2014 and Ph. D. degree in pattern recognition and intelligent systems from Shenyang Institute of Automation, Chinese Academy of Sciences in 2019. From 2017 to 2018, he visited the University of Arizona as a joint Ph. D. student. Now he works at School of Aeronautics and Astronautics, Central South University. His research interests include polarization imaging, image processing, image measurement, machine learning and pattern recognition.
\end{IEEEbiography}







\end{document}